\documentclass[preprint, amsmath, amssymb, aps, superscriptaddress, nofootinbib, nobibnotes, onecolumn]{revtex4}

\pdfoutput=1
\usepackage{graphicx}
\graphicspath{{figures/}}
\usepackage{amsmath}
\usepackage{amsfonts}
\usepackage{amssymb}
\usepackage{multirow}
\usepackage[colorlinks=true, linkcolor=red, citecolor=blue]{hyperref}
\usepackage[capitalise]{cleveref}
\crefname{figure}{Fig.}{Figs.}
\Crefname{figure}{Fig.}{Figs.}

\def\({\left(}
\def\){\right)}
\def\[{\left[}
\def\]{\right]}
\def\be{\begin{eqnarray}}
\def\ee{\end{eqnarray}}

\usepackage{acro}
\DeclareAcronym{BH}{
	short = BH ,
	long  = black hole
}
\DeclareAcronym{BBH}{
	short = BBH ,
	long  = binary black hole
}
\DeclareAcronym{GW}{
	short = GW ,
	long  = gravitational wave
}
\DeclareAcronym{GR}{
	short = GR ,
	long  = general relativity
}
\DeclareAcronym{MG}{
	short = MG ,
	long  = Modified Gravity
}
\DeclareAcronym{MDR}{
	short = MDR ,
	long  = modified dispersion relation
}
\DeclareAcronym{UV}{
	short = UV ,
	long  = ultraviolet
}
\DeclareAcronym{aLIGO}{
	short = aLIGO ,
	long = Advanced Laser Interferometer Gravitational-Wave Observatory
}
\DeclareAcronym{ET}{
	short = ET ,
	long  = Einstein Telescope
}
\DeclareAcronym{LISA}{
	short = LISA ,
	long  = Laser Interferometer Space Antenna
}
\DeclareAcronym{DECIGO}{
	short = DECIGO ,
	long  = Deci-Hertz Interferometer Gravitational Wave Observatory
}
\DeclareAcronym{QFT}{
	short = QFT ,
	long = quantum field theory
}
\DeclareAcronym{EFT}{
	short = EFT ,
	long = effective field theory
}
\DeclareAcronym{LCDM}{
	short = $\Lambda$CDM ,
	long = cosmological constant plus cold dark matter
}
\DeclareAcronym{SME}{
	short = SME ,
	long = Standard Model Extension
}
\DeclareAcronym{SNR}{
	short = SNR ,
	long = signal-to-noise ratio
}
\DeclareAcronym{rms}{
	short = rms ,
	long = root-mean-square
}
\DeclareAcronym{PSD}{
	short = PSD ,
	long = power spectral density
}
\DeclareAcronym{ppE}{
	short = ppE ,
	long = parameterized post-Einstein framework
}
\DeclareAcronym{LAL}{
	short = \texttt{LALSuite} ,
	long  = LSC Algorithm Library Suite
}
\DeclareAcronym{pN}{
	short = PN ,
	long  = post-Newtonian
}

\begin{document}

\title{Exploring the CPT violation and birefringence of gravitational waves with ground- and space-based gravitational-wave interferometers}

\author{Sai Wang}
\affiliation{Theoretical Physics Division, Institute of High Energy Physics, Chinese Academy of Sciences, Beijing 100049, People's Republic of China}

\begin{abstract}
\noindent
In the gravitational sector, we study the CPT violation and birefringence of gravitational waves. In presence of the CPT violation, a relative dephasing is generated between two circular polarization states of gravitational waves. This effect induces the birefringence of gravitational waves. We predict the gravitational waveform corrected by it and estimate the expected constraints on it from Advanced Laser Interferometer Gravitational-Wave Observatory, Einstein Telescope and Laser Interferometer Space Antenna. 
\end{abstract}

\maketitle
\acresetall

\section{Introduction}\label{sec:intro}
\noindent
In special relativity, the Lorentz symmetry is a fundamental invariance of physical laws in Minkowski spacetime. The CPT symmetry is also a fundamental invariance under the simultaneous transformations of charge conjugation (C), parity transformation (P), and time reversal (T). The CPT symmetry is exact in any Lorentz invariant local \acl{QFT} with a Hermitian Hamiltonian. Both Lorentz and CPT symmetries have been well tested by experiments of high energy physics (see review in Ref.~\cite{Kostelecky:2008ts}). For gravitational sector, the local Lorentz symmetry is one of the fundamental pillars of \ac{GR}. At extremely high energy scales, however, gravitational Lorentz symmetry is expected to be broken in theories of quantum gravity, such as deformed special relativity \cite{AmelinoCamelia:2000ge,AmelinoCamelia:2000mn,KowalskiGlikman:2001gp,Magueijo:2001cr}, Horava-Lifshitz gravity \cite{Horava:2009uw}, loop quantum gravity \cite{Gambini:1998it,Alfaro:2001rb}, non-commutative geometry \cite{Carroll:2001ws,Douglas:2001ba}, superstring theory \cite{Kostelecky:1988zi}, etc. The CPT symmetry may not hold any more in absence of local Lorentz symmetry, while the CPT violation necessarily violates Lorentz symmetry in an interacting theory \cite{Greenberg:2002uu}. The tests of CPT symmetry are therefore essential parts of testing Lorentz symmetry.
 
An \acl{EFT} of gravitational local Lorentz violation has been constructed in a framework of \ac{SME} \cite{Kostelecky:2003fs}. The local Lorentz-violating operators of arbitrary dimensions were introduced into gravitational action. In the limit of flat spacetime, a Lagrangian was expanded linearly around the Minkowski metric, and a covariant dispersion relation of \acp{GW} was deduced correspondingly \cite{Kostelecky:2016kfm}. 
The birefringence of \acp{GW} can be generated by the CPT-odd operators on which we focus. A dispersion relation is split into two branches which are related to the two circular polarization states, respectively. This implies that the two modes have a relative group velocity, which leads to an arrival-time difference between them in temporal domain. Or equivalently, there is a relative dephasing between the two modes in frequency domain. This phenomena seems like a correspondence to the birefringence of electromagnetic waves \cite{Myers:2003fd}. Therefore, one can test the CPT symmetry in gravitational sector by precisely measuring \acp{GW} which are emitted by compact binaries, for example, \acp{BBH}.

The recent discovery of \acp{GW} from compact binary coalescences, reported by \ac{aLIGO} and Virgo Collaborations \cite{Abbott:2016blz,Abbott:2016nmj,Abbott:2017vtc,Abbott:2017oio,TheLIGOScientific:2017qsa,Abbott:2017gyy,Monitor:2017mdv}, has opened an observational window to explore Lorentz and CPT violation in gravitational sector. For example, GW170817 has placed the most stringent constraints on the difference between the Lorentz-violating coefficients of mass dimension four in gravitational and photon sectors \cite{Monitor:2017mdv}. This existing study is based on multi-messenger measurements of a relative group velocity of \acp{GW} and electromagnetic waves. Due to an arrival-time difference between two circular polarizations in presence of the birefringence, there is a slight splitting of the peak at the maximal amplitude of the observed \ac{GW} signal. However, there are not significant evidences for such a splitting reported by \ac{aLIGO} \cite{Abbott:2016blz}, and hence an upper limit on the birefringence of \acp{GW} can be obtained by measuring the width of the peak. Based on the GW150914 signal in temporal domain, the upper limit of $\sim\mathcal{O}(10^{-14})\mathrm{m}$ has been placed on the dimension-five CPT-odd Lorentz-violating operators in gravitational sector \cite{Kostelecky:2016kfm}. 

In this work, we will study the constraints on the CPT symmetry breaking and birefringence in gravitational sector from three \ac{GW} experiments, including the second-generation ground-based \ac{aLIGO} \cite{TheLIGOScientific:2014jea}, the third-generation ground-based \ac{ET} \cite{Punturo:2010zz}, and the space-based \ac{LISA} \cite{Danzmann:1996da}. 
By introducing the CPT-violating dispersion in \ac{SME}, we will explore impacts of the birefringence on propagations of the two circular polarizations of \acp{GW}, which are emitted by distant astrophysical sources, and obtain corresponding modifications to \ac{GR} gravitational waveform by following Refs.~\cite{Mirshekari:2011yq,Yunes:2016jcc}. 
We will further use Fisher information matrix to estimate experimental sensitivities of these detectors to the CPT-violating parameter.
In particular, we expect to obtain some bounds on an effective characteristic length scale, below which the CPT violation in gravitational sector might emerge.

The rest of this paper is arranged as follows. In section \ref{sec:mdr}, we introduce the CPT-violating dispersion relation in gravitational sector, and study its modifications to \ac{GR} gravitational waveform. In section \ref{sec:fisher}, we introduce Fisher information matrix. In section \ref{sec:cptgw}, we show the expected constraints on the CPT-violating parameter and birefringence of \acp{GW}. The conclusion and discussion are summarized in section \ref{sec:conclu}.

\section{Deformations of gravitational waveform due to the CPT violation}\label{sec:mdr}
\noindent
We study the CPT violation and birefringence of \acp{GW} in the model-independent \ac{SME} \cite{Kostelecky:2016kfm}. The birefringence can be induced by the CPT-odd operators of dimension higher than four.
The CPT-violating dispersion influences the propagation of \acp{GW} from astrophysical sources to detectors \cite{Keppel:2010qu,Mirshekari:2011yq,Yunes:2016jcc}. In principle, it also contributes to the \ac{GW} emission process of the sources \cite{Bailey:2017lbo}. As argued in Ref.~\cite{Tso:2016mvv}, for a distant \ac{GW} source, the propagation effect can accumulate along the trajectory and could dominate over the emission effect. This is an ansatz in this study.
As mentioned above, due to the birefringence of \acp{GW}, the dispersion relation is split into two branches for the two circular polarizations, which thus take different propagating group velocities. Therefore, there is an arrival-time difference between the two modes, or equivalently the birefringence induces a relative dephasing between the two modes in frequency domain. 

We explore the birefringence of \acp{GW} in a phenomenological framework, which was followed by e.g. Refs.~\cite{Kostelecky:2016kfm,Tso:2016mvv,Yunes:2016jcc}.
The CPT-violating dispersion relation of \acp{GW} takes the following form
\begin{equation}\label{eq:mdr}
E^2=p^2\pm\zeta p^{\alpha}\ ,
\end{equation}
where a $\pm$ symbol reflects the presence of birefringence, and $\alpha$ denotes a dimensionless parameter. Here the left-handed circular polarization takes ``$+$'' while the right-handed one takes ``$-$'' in Eq.~(\ref{eq:mdr}). Throughout this paper, we appoint $G=c=1$ in which $G$ and $c$ denote Newton's constant and the speed of light, respectively. For any given $\zeta$, we define an effective length scale as 
$\ell_\alpha=h_{\mathrm{P}}|\zeta|^{\frac{1}{\alpha-2}}$,
below which the CPT violation could emerge.
Here $h_{\mathrm{P}}$ is Planck constant.
In this study, we focus on the dispersion relation (\ref{eq:mdr}) with odd index $\alpha=3,5,7,...$, which corresponds to a rotation-invariant limit of the leading-order dispersion in \ac{SME} (see Eq.~(11) in Ref.~\cite{Kostelecky:2016kfm}).

For a given dispersion relation of form $E^2=p^2+Ap^a$, in which $A$ and $a$ denote two Lorentz-violating parameters, the dephasing of \acp{GW} has been extensively studied recently \cite{Mirshekari:2011yq,Yunes:2016jcc}. In such a case, the speed of \acp{GW} would be different from the speed of light, which is exactly the speed of \acp{GW} in \ac{GR}. Compared to the \ac{GR} gravitational waveform, in frequency domain, the modified waveform thus obtains a dephasing which accumulates due to a distant propagation of \acp{GW}. In this work, we follow the same way and adopt the similar formulae, except that we introduce different dispersions to the two circular polarizations.
To be specific, we take the dispersion relation as $E^2=p^2+\zeta_{\mathrm{L,R}}p^{\alpha}$, where the subscripts $_\mathrm{L}$ and $_\mathrm{R}$ denote the left- and right-handed modes, respectively.
Due to $\zeta_{\mathrm{L}}\neq\zeta_{\mathrm{R}}$, the dephasing of left-handed mode is different from that of right-handed one. 
Therefore, between the two modes, there is a relative dephasing which can be tested by \ac{GW} detectors, as suggested by Ref.~\cite{Kostelecky:2016kfm}.
From Eq.~(\ref{eq:mdr}), we find $\zeta_{\mathrm{L}}=-\zeta_{\mathrm{R}}$, which implies opposite-sign dephasings for the two modes.

For each circular polarization, we can deduce modifications to its phase in the \ac{GR} waveform from the dispersion relation in Eq.~(\ref{eq:mdr}). The deduction process is as same as those followed by Refs.~\cite{Mirshekari:2011yq,Yunes:2016jcc}, except that we consider the circular polarizations here.
Denoting the \ac{GR} waveform with $h^{\mathrm{gr}}_{\mathrm{L,R}}$, we find that a \ac{GW} rotates along its propagating trajectory and arrives as $h_{\mathrm{L,R}}$ given by \cite{Tso:2016mvv}
\be
h_{\mathrm{L,R}}=h^{\mathrm{gr}}_{\mathrm{L,R}}~\mathrm{exp}\({\pm i\delta\Psi}\)\ ,
\ee
where a ``$\pm$'' symbol takes ``$+$'' for left-handed mode and ``$-$'' for right-handed one, and $\delta{\Psi}$ denotes a frequency-dependent phase deformation due to the CPT violation, i.e., 
\be\label{eq:dephasing}
\delta{\Psi}=
\frac{\lambda}{\alpha-1} \(\pi \mathcal{M}_{z} f\)^{\alpha-1}\ .
\ee 
The dephasing (\ref{eq:dephasing}) is as same as that in Ref.~\cite{Mirshekari:2011yq}, since we fix the index as $\alpha=3,5,7,...$.
Throughout this work, $m_i$ ($i=1, 2$) is an $i$-th component mass of a compact binary in the source frame, $M_z=(m_1+m_2)(1+z)$ a total mass of the binary in the observer frame,
$\eta=m_1m_2(m_1+m_2)^{-2}$ a symmetric mass ratio,  $\mathcal{M}_z=M_z\eta^{3/5}$ a chirp mass,
and $f$ an observed \ac{GW} frequency.
A parameter $\lambda$ is defined in terms of $\alpha$ and $\zeta$, i.e.,
\be\label{eq:zetatoxi}
\lambda=\frac{h_{\mathrm{P}}^{\alpha-2} \zeta}{\pi^{\alpha-2}{\mathcal{M}_z}^{\alpha-1}}\int_{0}^{z}\frac{(1+z^\prime)^{\alpha-2}dz^\prime}{H_0\sqrt{\Omega_m(1+z^\prime)^3+\Omega_\Lambda}}\ ,
\ee
where $z$ is a cosmological redshift to the source. 
Here we assume a spatially-flat \ac{LCDM} model, the independent parameters of which are fixed to their best-fit values from Planck 2015 results \cite{Ade:2015xua}, i.e., Hubble constant $H_0=67.74~\mathrm{km}~\mathrm{s}^{-1}~\mathrm{Mpc}^{-1}$, the fraction of matter density today $\Omega_m=0.3089$, and the fraction of dark energy density today $\Omega_\Lambda=1-\Omega_m$.

The circular polarization states are conventionally decomposed as the ``$+$'' and ``$\times$'' states, namely, $h_{\mathrm{L,R}}=h_{+}\pm i h_{\times}$ and $h^{\mathrm{gr}}_{\mathrm{L,R}}=h^{\mathrm{gr}}_{+}\pm i h^{\mathrm{gr}}_{\times}$.
Therefore, we obtain $h_{+,\times}$ as 
\be\label{eq:hp1}
h_{+}&=&\cos(\delta\Psi)h^{\mathrm{gr}}_{+}-\sin(\delta\Psi)h^{\mathrm{gr}}_{\times}\ ,\\
\label{eq:hc1}
h_{\times}&=&\sin(\delta\Psi)h^{\mathrm{gr}}_{+}+\cos(\delta\Psi)h^{\mathrm{gr}}_{\times}\ ,
\ee
where $h^{\mathrm{gr}}_{+,\times}$ denote gravitational waveform in \ac{GR}, i.e.,
\be
h^{\mathrm{gr}}_{+}&=&\frac{1}{2}(1+\cos^2\iota)\tilde{h}^{\mathrm{gr}}\ ,\\
h^{\mathrm{gr}}_{\times}&=&(i \cos\iota)\tilde{h}^{\mathrm{gr}}\ ,
\ee
where $\iota$ is an inclination angle of the binary, and $\tilde{h}^{\mathrm{gr}}$ is given by a non-spinning limit of \texttt{IMRPhenomB} \cite{Ajith:2009bn}, including the inspiral--merger--ringdown evolution of a binary coalescence. \texttt{IMRPhenomB} is available up to the asymmetric mass ratio of $4$.
Since we consider binaries with approximately equal component masses, it is enough for this purpose. Its explicit expression is listed in \ref{sec:waveform}.
In addition, in the limit of $|\delta\Psi|\ll1$, we expand $h_{+,\times}$ to linear order in $\delta\Psi$ and obtain $h_{+}=h_{+}^{\mathrm{gr}}-i\delta\Psi h_{\times}^{\mathrm{gr}}$ and $h_{\times}=h_{\times}^{\mathrm{gr}}+i\delta\Psi h_{+}^{\mathrm{gr}}$. This is so-called amplitude birefringence \cite{Yunes:2010yf}.

Given the waveform $h_{+,\times}$, one can define an ``observed'' waveform in frequency domain.
For an $i$-th interferometer of a given experiment, the ``observed'' waveform is written as \cite{Sathyaprakash:2009xs}
\be\label{eq:netwaveform}
{h}^{(i)}(f)=\epsilon F_{+}^{(i)}(\bar{\theta},\bar{\phi},\psi){h}_{+}(f)+\epsilon F_{\times}^{(i)}(\bar{\theta},\bar{\phi},\psi){h}_{\times}(f)\ ,
\ee
where $F_{+,\times}^{(i)}$ denote a set of two pattern functions, and one has $\epsilon=1$ for \ac{aLIGO} with arms at a $\pi/2$ opening angle \cite{Li:2013lza}, $\epsilon=\sqrt{3}/2$ for \ac{ET} with arms at a $\pi/3$ opening angle \cite{Zhao:2010sz,Li:2013lza}, and $\epsilon=\sqrt{3}/2$ for \ac{LISA} \cite{Klein:2015hvg,Babak:2017tow}.
For \ac{aLIGO}, we consider a single interferometer. One explicitly writes $F_{+,\times}^{(1)}$ as
\be
F_+^{(1)}&=&\frac{1}{2}\(1+\cos^2\bar{\theta}\)\cos2\bar{\phi}\cos2\psi-\cos\bar{\theta}\sin2\bar{\phi}\sin2\psi\ ,\\
F_{\times}^{(1)}&=&\frac{1}{2}\(1+\cos^2\bar{\theta}\)\cos2\bar{\phi}\sin2\psi-\cos\bar{\theta}\sin2\bar{\phi}\cos2\psi\ .
\ee
Here we have a source location (i.e., polar angle $\bar{\theta}$ and azimuthal angle $\bar{\phi}$) and a polarization angle (i.e., $\psi$) defined in the detector frame. 
For \ac{ET} \cite{Zhao:2010sz}, we consider three interferometers in total. The two additional sets of antenna pattern functions are given by 
\be
F^{(2)}_{+,\times}(\bar{\theta},\bar{\phi},\psi)&=&F^{(1)}_{+,\times}(\bar{\theta},\bar{\phi}+2\pi/3,\psi)\ ,\\
F^{(3)}_{+,\times}(\bar{\theta},\bar{\phi},\psi)&=&F^{(1)}_{+,\times}(\bar{\theta},\bar{\phi}+4\pi/3,\psi)\ .
\ee 
For \ac{LISA} \cite{Klein:2015hvg,Babak:2017tow}, we consider two interferometers in total. The second set of antenna pattern functions is given by \be
F^{(2)}_{+,\times}(\bar{\theta},\bar{\phi},\psi)=F^{(1)}_{+,\times}(\bar{\theta},\bar{\phi}-\pi/4,\psi)\ .
\ee 
When $\psi=\pi/8$ is set in the following, we can get a relation of $F_+^{(i)}=F_\times^{(i)}$. We only study the optimally-oriented compact binary coalescences, which are face-on and located directly above the detectors, i.e. $\iota=\bar{\theta}=0$. We can thus set $\bar{\phi}=0$ here.

\section{Fisher information matrix}\label{sec:fisher}
\noindent
Fisher information matrix \cite{Finn:1992xs,Cutler:1994ys,Poisson:1995ef} 
is performed to get the expected constraints on the CPT violation and on the birefringence of \acp{GW}, given a \ac{GW} detection which is consistent with \ac{GR}. Fisher matrix is defined as
\be\label{eq:fishermatrix}
F_{ab}=\sum_{i=1}^{n}\(\frac{\partial h^{(i)}}{\partial \theta_a}\bigg|\frac{\partial h^{(i)}}{\partial \theta_b}\)\ ,
\ee
where $\theta_a$ denotes $a$-th parameter, $n$ is a total number of interferometers for a given experiment, and we define an inner product between two waveforms $\tilde{h}_1$ and $\tilde{h}_2$ as 
\be\label{eq:innerproduct}
\(h_1|h_2\)=2\int_{f_\mathrm{low}}^{f_{\mathrm{high}}}\frac{{h}_1^\ast(f){h}_2(f)+{h}_1(f){h}_2^\ast(f)}{S_h(f)}df\ ,
\ee
where a $^\ast$ symbol is a complex conjugate, and $S_h(f)$ denotes a noise \ac{PSD} of the given detector. 
In \ref{sec:psd}, we summarize the noise \acp{PSD} of 
\ac{aLIGO} \cite{Ajith:2011ec}, \ac{ET} \cite{Mishra:2010tp}, 
and \ac{LISA} \cite{Klein:2015hvg,Babak:2017tow}. 
Here $f_{\mathrm{low}}$ is a lower-cutoff frequency of the detector, while $f_{\mathrm{high}}$ is an upper-cutoff frequency, at which the \ac{GW} detection terminates. 
\ac{aLIGO}, \ac{ET}, and \ac{LISA} are sensitive to frequency ranges $10-10^{4}\mathrm{Hz}$, $1-10^{4}\mathrm{Hz}$, and $10^{-4}-10^{-1}\mathrm{Hz}$, respectively.
In addition, a \ac{SNR} is defined as $\mathrm{SNR}=\sum_{i=1}^{n}\sqrt{(h^{(i)}(f)|h^{(i)}(f))}$.

A \ac{rms} uncertainty on $\theta_a$ is defined as a diagonal component of covariance matrix $C_{ab}$, i.e. $\Delta\theta_a=\sqrt{C_{aa}}$. 
Cramer-Rao bound \cite{cramer1999mathematical,Rao:1945} says an inequality of form $C\ge F^{-1}$. 
Once Fisher matrix is obtained, one uses Cholesky decomposition to get an inverse of Fisher matrix.
Therefore, a minimal uncertainty on $\theta_a$ is given by
\be\label{eq:error}
\Delta\theta_a=\sqrt{\(F^{-1}\)_{aa}}\ ,
\ee
which is determined by $a$-th diagonal component of the inverse of Fisher matrix.
Furthermore, one defines a normalized covariance matrix, i.e.,
$c_{ab}={\(F^{-1}\)_{ab}}/({\Delta\theta_{a}\Delta\theta_{b}})$,
to describe a cross correlation between $\theta_a$ and $\theta_b$.

We consider the optimally-oriented sources to estimate the constraints on the CPT-violating parameters. 
In frequency domain, the CPT violation induces the relative dephasing between the two circular polarizations of \acp{GW}.
It is enough for a single detector to measure such a relative quantity.
So we do not consider a network of \ac{GW} detectors. 
However, one should note that a full multi-detector Bayesian analysis, that simultaneously disentangle polarizations and fit for the polarization-dependent dispersion parameters, may achieve better constraints than those obtained just from looking at the waveform dephasing in a single detector. We leave such a detailed analysis to future works.
In this study, our parameter space is spanned by six dimensionless parameters, i.e.,
\be
\boldsymbol{\theta}=\{\ln d_{\mathrm{L}},\ln \mathcal{M},\ln\eta,\phi_c,f_0t_c,\lambda\}\ , 
\ee
where we use the chirp mass in the source frame, i.e. $\mathcal{M}=\mathcal{M}_z(1+z)^{-1}$, and $f_0$ denotes a characteristic frequency of the detector, typically chosen as a ``knee'' frequency or a frequency that makes the noise \ac{PSD} minimal.
In a fiducial model, we let $\phi_c=f_{0}t_c=\lambda=0$,
and let $d_{\mathrm{L}}\simeq1~\mathrm{Gpc}$ for \ac{aLIGO} and \ac{ET} while  $d_{\mathrm{L}}\simeq3~\mathrm{Gpc}$ for \ac{LISA}.
We consider $\mathcal{M}$ varying within $20-10^{3}M_{\odot}$ for \ac{aLIGO}, $20-3\times10^{3}M_{\odot}$ for \ac{ET}, and $10^{5}-10^{7}M_{\odot}$ for \ac{LISA}.
In addition, we study the dependence of our results on $\eta$, which varies from $0.16$ to $0.25$. 
For a given $\alpha$, once a constraint on $\lambda$ is obtained, we can deduce a constraint on $\zeta$ according to Eq.~(\ref{eq:zetatoxi}) and hence on the effective length scale $\ell_\alpha$.

\section{Constraints on the CPT violation and GW birefringence from gravitational-wave interferometers}\label{sec:cptgw}
\begin{figure}[htbp]
\includegraphics[width=1\columnwidth]{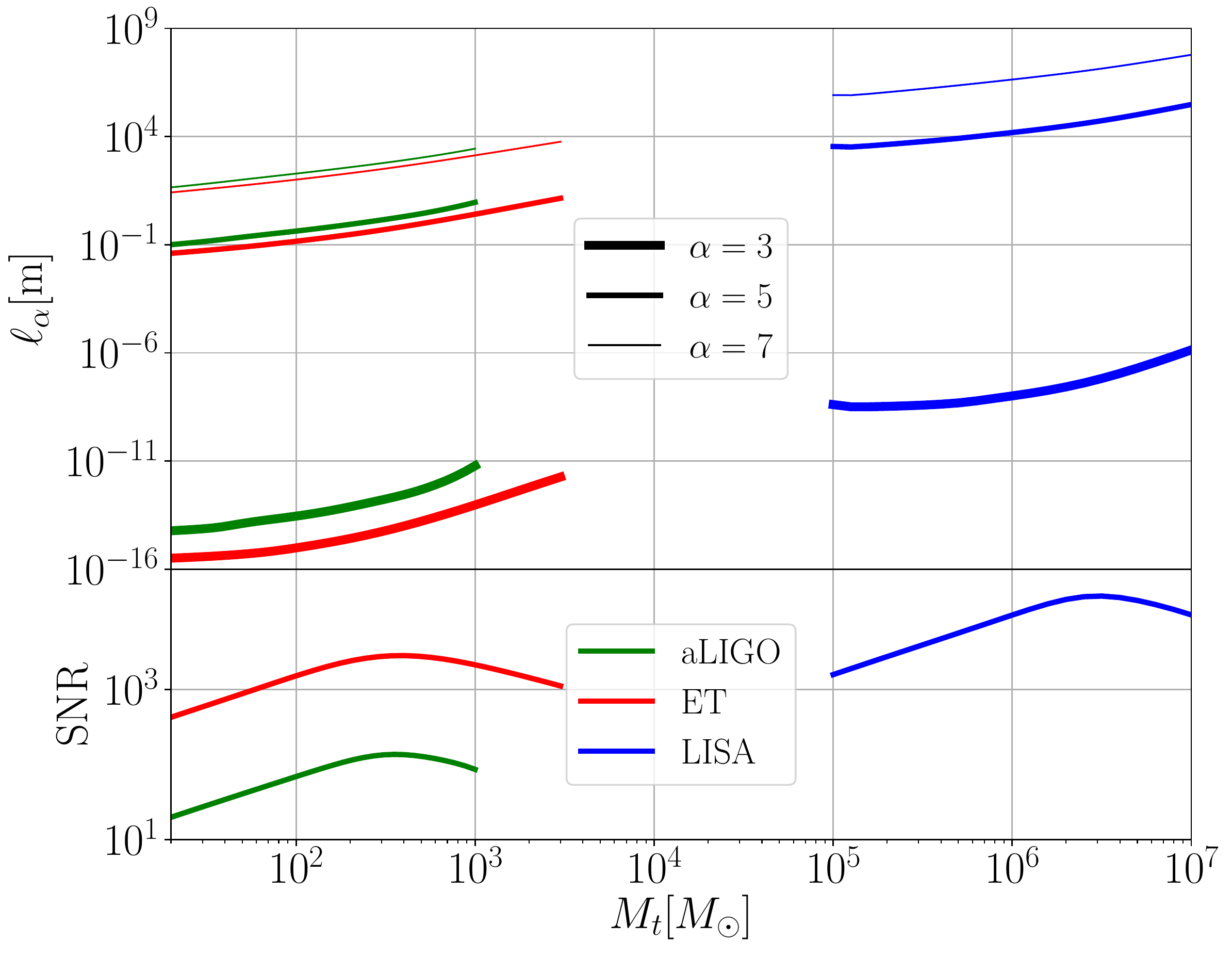}
\caption{The upper limits on the CPT-violating parameter $\ell_{\alpha}$ ($\alpha=3,5,7$) from \ac{aLIGO} (green), \ac{ET} (red), and \ac{LISA} (blue). The \acp{SNR} are also obtained for the three detectors here. }
\label{fig:figureerror}
\end{figure}

\noindent
For a given $\alpha$, by using Fisher matrix, we obtain the $1\sigma$ uncertainty on $\lambda$ and then obtain an upper limit on $\ell_{\alpha}$. 
For the coalescing \acp{BBH} with equal component masses, Figure~\ref{fig:figureerror} shows such upper limits on $\ell_\alpha~(\alpha=3,5,7)$ expected from \ac{aLIGO} (green curve), \ac{ET} (red curve), and \ac{LISA} (blue curve).
The \acp{SNR} are also depicted for all the three detectors. Here $M_{t}=m_{1}+m_{2}$ is a total mass of the binary in the source frame. $\ell_{\alpha}$ is in units of $m$. We use thicker curves to denote lower-order $\alpha$ in the figure. 

We notice several generic characters about our results. For all the three detectors, the upper limits on $\ell_\alpha$ become higher with the increase of $\alpha$. 
This means that we obtain less stringent constraints on $\ell_\alpha$ for higher-order $\alpha$.
For a given $\alpha$, \ac{ET} can always has a higher sensitivity than \ac{aLIGO} in the same mass range.
However, for a higher-order $\alpha$, the difference in the sensitivities of \ac{ET} and \ac{aLIGO} becomes less significant. 
For equally spaced $\alpha$, e.g. $\alpha_{i}$ (i=1,2,3) with $\alpha_1<\alpha_2<\alpha_3$ and $2\alpha_2=\alpha_1+\alpha_3$, the ratio $\ell_{\alpha_2}/\ell_{\alpha_1}$ is always larger than the ratio $\ell_{\alpha_3}/\ell_{\alpha_2}$.
The above general predictions could be explained as follows.
Based on Eq.~(\ref{eq:dephasing}), a higher-order $\alpha$ can generate a larger phase correction at higher frequencies. 
Naively, a higher-order $\alpha$ could be corresponded to a higher \ac{pN} order.
By contrast, the \ac{GW} detectors can measure the higher-order phase deformations at a worse level \cite{Yunes:2009ke,Arun:2006hn,Agathos:2013upa}. 
Obviously, our results arise from a balance between the above two competitive ingredients.
In addition, \ac{LISA} has a worse sensitivity, partly because the birefringence is proportional to $f^{\alpha-1}$ in Eq.~(\ref{eq:dephasing}).
To get a better knowledge of the above discussions, we estimate the orders of magnitude for the following typical cases.
For the chirp mass $\simeq30M_\odot$, \ac{ET} will reach the sensitivities $\ell_3\sim\mathcal{O}(10^{-16})\mathrm{m}$, $\ell_5\sim\mathcal{O}(10^{-2})\mathrm{m}$ and $\ell_7\sim\mathcal{O}(10)\mathrm{m}$, while \ac{aLIGO} will reach $\ell_3\sim\mathcal{O}(10^{-15})\mathrm{m}$, $\ell_5\sim\mathcal{O}(10^{-1})\mathrm{m}$ and $\ell_7\sim\mathcal{O}(10)\mathrm{m}$.
For the chirp mass $\simeq10^5M_\odot$, \ac{LISA} will reach $\ell_3\sim\mathcal{O}(10^{-9})\mathrm{m}$, $\ell_5\sim\mathcal{O}(10^{3})\mathrm{m}$ and $\ell_7\sim\mathcal{O}(10^{5})\mathrm{m}$.

For the coalescing \acp{BBH} with different component masses, e.g. $\eta=0.16,0.25$, Table~\ref{tab:tableerror} shows the \ac{rms} uncertainties on the source parameters (i.e. $\mathcal{M}$, $\eta$) and the CPT-violating parameter (i.e. $\lambda$), as well as their normalized cross-correlation coefficients, in the case of $\alpha=3$.
The uncertainty on $\lambda$ is converted to the upper limit on $\ell_3$ in this table.
Given the \ac{GW} signal, which is consistent with \ac{GR}, one would expect $\lambda$ to be smaller than the magnitudes in the sixth column of Table~\ref{tab:tableerror}. Therefore, the lower bound on $\ell_{3}$ is listed in the seventh column. 
To show the degeneracy of parameters, we show the cross-correlations among $\ln\mathcal{M}$, $\ln\eta$ and $\lambda$ in the last three columns.
\begin{table*}[!htp]
\centering
\renewcommand{\arraystretch}{1}
\begin{tabular}{| l | l | l | l | l | l | l | l | l | l |}
\hline
 & $M_{t}/M_\odot$ & $\eta$ & $\Delta\ln\mathcal{M}$ & $\Delta\ln\eta$ & $\Delta\lambda$ & $\ell_3/m$ & $c_{\mathcal{M}\eta}$ & $c_{\mathcal{M}\lambda}$ & $c_{\eta\lambda}$ \\
\hline
\multirow{4}{0.75cm}{aLIGO}
& 30 & 0.16 & $1.1\times10^{-3}$ & $1.5\times10^{-2}$ & $560.7$ & $1.9\times10^{-14}$ & $0.936$ & $0.566$ & $0.735$ \\
& 30 & 0.25 & $1.5\times10^{-3}$ & $2.2\times10^{-2}$ & $123.1$ & $7.3\times10^{-15}$ & $0.954$ & $0.293$ & $0.180$ \\
& 300 & 0.16 & $4.4\times10^{-2}$ & $5.6\times10^{-2}$ & $81.4$ & $2.8\times10^{-13}$ & $0.992$ & $-0.130$ & $-0.094$ \\
& 300 & 0.25 & $1.5\times10^{-2}$ & $1.4\times10^{-2}$ & $26.6$ & $1.6\times10^{-13}$ & $0.982$ & $-0.618$ & $-0.621$ \\
\hline
\multirow{6}{0.75cm}{ET}
&$30$&$	0.16$&$1.3\times10^{-5}$&$3.8\times10^{-4}$&$25.6$&$8.9\times10^{-16}$&$0.919$&$0.616$&$0.747$\\
&$30$&$	0.25$&$6.6\times10^{-6}$&$2.8\times10^{-4}$&$6.3$&$3.7\times10^{-16}$&$0.314$&$-0.076$&$0.448$\\
&$300$&$	0.16$&$8.8\times10^{-5}$&$5.1\times10^{-4}	$&$3.7$&$1.3\times10^{-14}	$&$0.822	$&$0.199$&$	0.467$\\
&$300$&$	0.25$&$6.6\times10^{-5}$&$1.5\times10^{-4}$&$0.9$&$5.4\times10^{-15}$&$	0.522	$&$-0.372$&$-0.265$\\
&$3000	$&$0.16$&$1.0\times10^{-2}$&$1.3\times10^{-2}$&$13.3$&$4.6\times10^{-12}	$&$0.998$&$0.546$&$0.557$\\
&$3000	$&$0.25$&$3.0\times10^{-3}$&$2.8\times10^{-3}$&$3.1$&$1.8\times10^{-12}	$&$0.996$&$-0.455$&$-0.455$\\
\hline
\multirow{6}{0.75cm}{LISA}
&$10^{5}$&$0.16$&$	4.7\times10^{-6}$&$	1.8\times10^{-4}$&$	34.6$&$	7.9\times10^{-9}$&$0.990$&$0.774$&$0.806$\\
&$10^{5}$&$	0.25$&$	8.5\times10^{-7}$&$6.8\times10^{-5}$&$10.4$&$4.0\times10^{-9}	$&$-0.244$&$-0.368$&$0.604$\\
&$10^{6}$&$	0.16$&$1.1\times10^{-5}$&$9.3\times10^{-5}$&$1.1$&$2.4\times10^{-8}	$&$0.846$&$0.369$&$0.628$\\
&$10^{6}$&$0.25	$&$8.8\times10^{-6}$&$6.7\times10^{-5}$&$0.3$&$1.0\times10^{-8}	$&$0.589$&$-0.474$&$-0.363$\\
&$10^{7}$&$0.16	$&$3.0\times10^{-4}$&$4.1\times10^{-4}$&$1.3$&$2.9\times10^{-6}$&$0.974$&$-0.055$&$0.002$\\
&$10^{7}$&$0.25	$&$2.0\times10^{-4}$&$1.8\times10^{-4}$&$0.4$&$1.4\times10^{-6}$&$0.988$&$-0.421$&$-0.420$\\
\hline
\end{tabular}
\caption{In the case of $\alpha=3$, for different value of $\eta$, the \ac{rms} uncertainties on $\ln\mathcal{M}$, $\ln\eta$ and $\lambda$, as well as their cross correlations. The uncertainties on $\ell_3$ are also listed here.} 
\label{tab:tableerror}
\end{table*}

According to Table~\ref{tab:tableerror}, we find that the symmetric mass ratio influences the constraints on the CPT violation and birefringence of \acp{GW}. Given $M_{t}$, the constraint on the CPT violation becomes more stringent with the increase of $\eta$, due to an increase of \ac{SNR}. Given $\eta_1$ and $\eta_2$ with $\eta_1<\eta_2$, the ratio $\ell_{\alpha}(\eta_2)/\ell_{\alpha}(\eta_1)$ becomes less significant with the increase of $M_{t}$, due to a smaller number of frequency modes in the sensitive range of the detectors. Therefore, it is helpful to use a compact binary system with equal component masses in the study of CPT violation and birefringence.
We also extend the parameter space to include $\iota$, $\psi$, $\bar{\theta}$ and $\bar{\phi}$, and vary their values around the fiducial values. Obviously, these additional parameters make the $\lambda$ constraints less stringent. However, the modifications are less than or around one order of magnitude in the most sensitive mass range, since we only utilize the relative dephasing between the two circular polarizations of \acp{GW}. In future works, we will study the network of \ac{GW} interferometers, which is expected to reduce the uncertainties \cite{Schutz:2011tw}.

\section{Conclusions and Discussion}\label{sec:conclu}
\noindent
In this paper, we explored the influences on the propagation of \acp{GW} from the CPT violation and birefringence in the gravitational sector \cite{Kostelecky:2016kfm,Tso:2016mvv,Yunes:2016jcc}. We found a relative dephasing between the two circular polarization modes of \acp{GW}, and obtained the corresponding gravitational waveform, which is corrected by the CPT-violating parameter. Considering distant compact binary coalescences, we estimated the projected constraints on the CPT-violating parameter from \ac{aLIGO}, \ac{ET}, and \ac{LISA}. Among these experiments, we expect \ac{ET} to be most sensitive to the CPT violation and birefringence. This study involved only the relative dephasing between the two circular polarization states and was therefore model-independent. Future \ac{GW} detections are expected to shed light on the CPT violation and birefringence in the gravitational sector. If such deviations from \ac{GR} were detected, we can use them to infer the degree of CPT violation and birefringence. If not, we can constrain the magnitude of the CPT-violating parameter to some interesting levels.

We made several approximations which are worth being revisited in the future. For example, we only considered the kinematical propagation of \acp{GW} while disregarded the dynamical generation process, which may generate some corrections to our results. For gravitational waveform, we did not take into account additional parameters such as eccentricity, spins, precession, etc. By contrast, it has been found that the spins in binaries can worsen the projected constraints on Lorentz violation \cite{Samajdar:2017mka}. In addition, we focus on the isotropic limit for the leading-order dispersion relation of \acp{GW} in \ac{SME}. In fact, there are a lot of anisotropic coefficients which can also lead to the birefringence. For the above concerns, we leave detailed analysis to future works. 
Furthermore, the \ac{GR} waveform in Eq.~(\ref{eq:waveform_sai}) is available up to the 3.5\ac{pN} order, which is lower order than the contributions from the CPT violation. This means that all the degeneracies between these higher-\ac{pN}-order terms and the CPT-violating terms are neglected. This is just feasible if we were only interested in how well the birefringence can be constrained by observations.

\begin{acknowledgements}\noindent
This work is supported in part by a grant from the Institute of High Energy Physics, Chinese Academy of Sciences. The author appreciates Prof.~T.G.F.~Li, Dr.~Y.-F.~Wang and Dr.~Z.-C.~Zhao for helpful discussions on this project and also Prof.~A.~Koskelety and Prof.~Q.~Bailey for useful comments on our manuscript of early version. 
\end{acknowledgements}

\appendix
\section{The \texttt{IMRPhenomB} waveform}\label{sec:waveform}
\noindent
In the non-spinning limit, the \texttt{IMRPhenomB} waveform \cite{Ajith:2009bn} takes the following form
\be\label{eq:waveform_sai}
\tilde{h}^{\mathrm{gr}}(f)=\frac{1}{d_{\mathrm{L}}}\mathcal{B}(f;M_z,\eta)~\mathrm{exp}\({i\Psi(f;M_z,\eta,\phi_c,t_c)}\)\ ,
\ee
where $\phi_{c}$ and $t_{c}$ denote a coalescence phase and time, respectively.
A luminosity distance to redshift $z$ is given by 
\be
 d_L=\frac{1+z}{H_0}\int_{0}^{z}\frac{dz^\prime}{\sqrt{\Omega_m(1+z^\prime)^3+\Omega_\Lambda}}\ .
\ee
Here $\mathcal{B}$ is explicitly expressed as
\be
\mathcal{B}(f)=\(\frac{5\eta}{24}\)^{1/2}\frac{M^{5/6}_z}{\pi^{2/3}f_{1}^{7/6}}
\begin{cases}
f^{\prime -7/6}\(1+\sum_{i=2}^{3}\alpha_i v^i\)~~~~~~~~\mathrm{for}~f<f_1 \\
w_m f^{\prime -2/3}\(1+\sum_{i=1}^{2}\epsilon_i v^i\)~~~~\mathrm{for}~f_1\le f<f_2\\
w_r\frac{\sigma/2\pi}{(f-f_2)^2+\sigma^2/4}~~~~~~~~~~~~~~~~~~~\mathrm{for}~f_2\le f\le f_3\\
0~~~~~~~~~~~~~~~~~~~~~~~~~~~~~~~~~~~~~~~~~\mathrm{for}~f>f_3
\end{cases}
\ee
where one defines two dimensionless parameters $f^\prime=f/f_1$ and $v = (\pi M_z f)^{1/3}$ for simplicity, and $w_m$ and $w_r$ are two normalization constants that make $\mathcal{B}$ continuous.
The parameters $\alpha_i$, $\epsilon_i$, $f_i$, and $\sigma$ are expressed in terms of $M_z$ and $\eta$, namely,
\be
&&\alpha_2=-323/224+451\eta/168\ ,~~~~
\alpha_3=0\ ,\\
&&\epsilon_1=-1.8897\ ,~~~~
\epsilon_2=1.6557\ ,\\
&&\pi M_z f_1=(1-4.455+3.521)+0.6437\eta-0.05822\eta^2-7.092\eta^3\ ,\\
&&\pi M_z f_2 = (1-0.63)/2+0.1469\eta-0.0249\eta^2+2.325\eta^3\ , \\
&&\pi M_z \sigma = (1-0.63)/4 -0.4098\eta +1.829\eta^2-2.87\eta^3 \ ,\\
&&\pi M_z f_3 = 0.3236 -0.1331\eta -0.2714\eta^2 +4.922\eta^3\ .
\ee
The phase $\Psi$ is explicitly expressed as
\be\label{GRphase}
\Psi(f)=2\pi f t_c+\phi_c+\frac{3}{128\eta v^5}\(1+\sum_{k=2}^{7}v^{k}\psi_{k}\)\ ,
\ee
where the coefficients $\psi_k$ are expressed in terms of $\eta$, i.e.
\be
&&\psi_2=3715/756-920.9\eta+6742\eta^2-1.34\times10^4\eta^3\ ,\\
&&\psi_3=-16\pi+1.702\times10^4\eta-1.214\times10^5\eta^2+2.386\times10^5\eta^3\ ,\\
&&\psi_4=15293365/508032-1.254\times10^5\eta+8.735\times10^5\eta^2-1.694\times10^6\eta^3\ ,\\
&&\psi_6=0-8.898\times10^5\eta+5.981\times10^6\eta^2-1.128\times10^7\eta^3\ ,\\
&&\psi_7=0+8.696\times10^5\eta-5.838\times10^6\eta^2+1.089\times10^7\eta^3\ .
\ee

\section{The noise PSDs of gravitational-wave detectors}\label{sec:psd}
\noindent
For the designed sensitivity of \ac{aLIGO} \cite{Thrane:2013oya}, we use the noise \ac{PSD} as \cite{Ajith:2011ec}
\be
S_h(f)=10^{-48}\(0.0152x^{-4}+0.2935x^{9/4}+2.7951x^{3/2}-6.5080x^{3/4}+17.7622\)~\mathrm{Hz}^{-1}\ ,\nonumber\\
\ee
\noindent
where $x=f/245.4\mathrm{Hz}$.
For \ac{ET}, we use the noise \ac{PSD} as \cite{Mishra:2010tp} 
\be
S_{h}(f)=10^{-50}\(2.39\times 10^{-27}x^{-15.64} +0.349 x^{-2.145} +1.76 x^{-0.12} +0.409 x^{1.1} \)^{2} ~\mathrm{Hz}^{-1}\ ,\nonumber\\
\ee
\noindent
where $x=f/100\mathrm{Hz}$.
For \ac{LISA}, we use the noise \ac{PSD} as \cite{Klein:2015hvg,Babak:2017tow}
\be
S_h(f)=\frac{20}{3}\(\frac{4S_n^{\mathrm{acc}}+2S_n^{\mathrm{loc}}+S_n^{\mathrm{sn}}+S_n^{\mathrm{omn}}}{L^2}\)\[1+\(\frac{2Lf}{0.41}\)^2\]~\mathrm{Hz}^{-1}\ ,
\ee
\noindent
where $L=2.5\times 10^9 \mathrm{m}$ is the length of arm, and  one has the following expressions
\be
&&S_n^{\mathrm{acc}}=\(\frac{1\mathrm{Hz}}{2\pi f}\)^{4}\[ 9\times 10^{-30}+3.24\times 10^{-28}\[\(\frac{3\times 10^{-5}\mathrm{Hz}}{f}\)^{10} +\(\frac{10^{-4}\mathrm{Hz}}{f}\)^{2}\]\]~\mathrm{m}^2\mathrm{Hz}^{-1}\ , \nonumber\\
&&\\
&&S_n^{\mathrm{loc}}=2.89\times 10^{-24}~\mathrm{m}^2\mathrm{Hz}^{-1}\ , \\ 
&&S_n^{\mathrm{sn}}=7.92\times 10^{-23}~\mathrm{m}^2\mathrm{Hz}^{-1}\ , \\
&&S_n^{\mathrm{omn}}=4.00\times 10^{-24}~\mathrm{m}^2\mathrm{Hz}^{-1}\ ,
\ee
\noindent
where $S_n^{\mathrm{acc}}$, $S_n^{\mathrm{loc}}$, $S_n^{\mathrm{sn}}$ and $S_n^{\mathrm{omn}}$ denote noises due to the low-frequency acceleration, local interferometer noise, shot noise and other measurement noise, respectively.

\vspace{0.1cm}

\bibliography{gw-cpt}

\begin{thebibliography}{51}
\expandafter\ifx\csname natexlab\endcsname\relax\def\natexlab#1{#1}\fi
\expandafter\ifx\csname bibnamefont\endcsname\relax
  \def\bibnamefont#1{#1}\fi
\expandafter\ifx\csname bibfnamefont\endcsname\relax
  \def\bibfnamefont#1{#1}\fi
\expandafter\ifx\csname citenamefont\endcsname\relax
  \def\citenamefont#1{#1}\fi
\expandafter\ifx\csname url\endcsname\relax
  \def\url#1{\texttt{#1}}\fi
\expandafter\ifx\csname urlprefix\endcsname\relax\def\urlprefix{URL }\fi
\providecommand{\bibinfo}[2]{#2}
\providecommand{\eprint}[2][]{\url{#2}}

\bibitem[{\citenamefont{Kostelecky and Russell}(2011)}]{Kostelecky:2008ts}
\bibinfo{author}{\bibfnamefont{V.~A.} \bibnamefont{Kostelecky}}
  \bibnamefont{and} \bibinfo{author}{\bibfnamefont{N.}~\bibnamefont{Russell}},
  \bibinfo{journal}{Rev. Mod. Phys.} \textbf{\bibinfo{volume}{83}},
  \bibinfo{pages}{11} (\bibinfo{year}{2011}), \eprint{0801.0287}.

\bibitem[{\citenamefont{Amelino-Camelia}(2001)}]{AmelinoCamelia:2000ge}
\bibinfo{author}{\bibfnamefont{G.}~\bibnamefont{Amelino-Camelia}},
  \bibinfo{journal}{Phys. Lett.} \textbf{\bibinfo{volume}{B510}},
  \bibinfo{pages}{255} (\bibinfo{year}{2001}), \eprint{hep-th/0012238}.

\bibitem[{\citenamefont{Amelino-Camelia}(2002)}]{AmelinoCamelia:2000mn}
\bibinfo{author}{\bibfnamefont{G.}~\bibnamefont{Amelino-Camelia}},
  \bibinfo{journal}{Int. J. Mod. Phys.} \textbf{\bibinfo{volume}{D11}},
  \bibinfo{pages}{35} (\bibinfo{year}{2002}), \eprint{gr-qc/0012051}.

\bibitem[{\citenamefont{Kowalski-Glikman}(2001)}]{KowalskiGlikman:2001gp}
\bibinfo{author}{\bibfnamefont{J.}~\bibnamefont{Kowalski-Glikman}},
  \bibinfo{journal}{Phys. Lett.} \textbf{\bibinfo{volume}{A286}},
  \bibinfo{pages}{391} (\bibinfo{year}{2001}), \eprint{hep-th/0102098}.

\bibitem[{\citenamefont{Magueijo and Smolin}(2002)}]{Magueijo:2001cr}
\bibinfo{author}{\bibfnamefont{J.}~\bibnamefont{Magueijo}} \bibnamefont{and}
  \bibinfo{author}{\bibfnamefont{L.}~\bibnamefont{Smolin}},
  \bibinfo{journal}{Phys. Rev. Lett.} \textbf{\bibinfo{volume}{88}},
  \bibinfo{pages}{190403} (\bibinfo{year}{2002}), \eprint{hep-th/0112090}.

\bibitem[{\citenamefont{Horava}(2009)}]{Horava:2009uw}
\bibinfo{author}{\bibfnamefont{P.}~\bibnamefont{Horava}},
  \bibinfo{journal}{Phys. Rev.} \textbf{\bibinfo{volume}{D79}},
  \bibinfo{pages}{084008} (\bibinfo{year}{2009}), \eprint{0901.3775}.

\bibitem[{\citenamefont{Gambini and Pullin}(1999)}]{Gambini:1998it}
\bibinfo{author}{\bibfnamefont{R.}~\bibnamefont{Gambini}} \bibnamefont{and}
  \bibinfo{author}{\bibfnamefont{J.}~\bibnamefont{Pullin}},
  \bibinfo{journal}{Phys. Rev.} \textbf{\bibinfo{volume}{D59}},
  \bibinfo{pages}{124021} (\bibinfo{year}{1999}), \eprint{gr-qc/9809038}.

\bibitem[{\citenamefont{Alfaro et~al.}(2002)\citenamefont{Alfaro,
  Morales-Tecotl, and Urrutia}}]{Alfaro:2001rb}
\bibinfo{author}{\bibfnamefont{J.}~\bibnamefont{Alfaro}},
  \bibinfo{author}{\bibfnamefont{H.~A.} \bibnamefont{Morales-Tecotl}},
  \bibnamefont{and} \bibinfo{author}{\bibfnamefont{L.~F.}
  \bibnamefont{Urrutia}}, \bibinfo{journal}{Phys. Rev.}
  \textbf{\bibinfo{volume}{D65}}, \bibinfo{pages}{103509}
  (\bibinfo{year}{2002}), \eprint{hep-th/0108061}.

\bibitem[{\citenamefont{Carroll et~al.}(2001)\citenamefont{Carroll, Harvey,
  Kostelecky, Lane, and Okamoto}}]{Carroll:2001ws}
\bibinfo{author}{\bibfnamefont{S.~M.} \bibnamefont{Carroll}},
  \bibinfo{author}{\bibfnamefont{J.~A.} \bibnamefont{Harvey}},
  \bibinfo{author}{\bibfnamefont{V.~A.} \bibnamefont{Kostelecky}},
  \bibinfo{author}{\bibfnamefont{C.~D.} \bibnamefont{Lane}}, \bibnamefont{and}
  \bibinfo{author}{\bibfnamefont{T.}~\bibnamefont{Okamoto}},
  \bibinfo{journal}{Phys. Rev. Lett.} \textbf{\bibinfo{volume}{87}},
  \bibinfo{pages}{141601} (\bibinfo{year}{2001}), \eprint{hep-th/0105082}.

\bibitem[{\citenamefont{Douglas and Nekrasov}(2001)}]{Douglas:2001ba}
\bibinfo{author}{\bibfnamefont{M.~R.} \bibnamefont{Douglas}} \bibnamefont{and}
  \bibinfo{author}{\bibfnamefont{N.~A.} \bibnamefont{Nekrasov}},
  \bibinfo{journal}{Rev. Mod. Phys.} \textbf{\bibinfo{volume}{73}},
  \bibinfo{pages}{977} (\bibinfo{year}{2001}), \eprint{hep-th/0106048}.

\bibitem[{\citenamefont{Kostelecky and Samuel}(1989)}]{Kostelecky:1988zi}
\bibinfo{author}{\bibfnamefont{V.~A.} \bibnamefont{Kostelecky}}
  \bibnamefont{and} \bibinfo{author}{\bibfnamefont{S.}~\bibnamefont{Samuel}},
  \bibinfo{journal}{Phys. Rev.} \textbf{\bibinfo{volume}{D39}},
  \bibinfo{pages}{683} (\bibinfo{year}{1989}).

\bibitem[{\citenamefont{Greenberg}(2002)}]{Greenberg:2002uu}
\bibinfo{author}{\bibfnamefont{O.~W.} \bibnamefont{Greenberg}},
  \bibinfo{journal}{Phys. Rev. Lett.} \textbf{\bibinfo{volume}{89}},
  \bibinfo{pages}{231602} (\bibinfo{year}{2002}), \eprint{hep-ph/0201258}.

\bibitem[{\citenamefont{Kostelecky}(2004)}]{Kostelecky:2003fs}
\bibinfo{author}{\bibfnamefont{V.~A.} \bibnamefont{Kostelecky}},
  \bibinfo{journal}{Phys. Rev.} \textbf{\bibinfo{volume}{D69}},
  \bibinfo{pages}{105009} (\bibinfo{year}{2004}), \eprint{hep-th/0312310}.

\bibitem[{\citenamefont{Kostelecky and Mewes}(2016)}]{Kostelecky:2016kfm}
\bibinfo{author}{\bibfnamefont{V.~A.} \bibnamefont{Kostelecky}}
  \bibnamefont{and} \bibinfo{author}{\bibfnamefont{M.}~\bibnamefont{Mewes}},
  \bibinfo{journal}{Phys. Lett.} \textbf{\bibinfo{volume}{B757}},
  \bibinfo{pages}{510} (\bibinfo{year}{2016}), \eprint{1602.04782}.

\bibitem[{\citenamefont{Myers and Pospelov}(2003)}]{Myers:2003fd}
\bibinfo{author}{\bibfnamefont{R.~C.} \bibnamefont{Myers}} \bibnamefont{and}
  \bibinfo{author}{\bibfnamefont{M.}~\bibnamefont{Pospelov}},
  \bibinfo{journal}{Phys. Rev. Lett.} \textbf{\bibinfo{volume}{90}},
  \bibinfo{pages}{211601} (\bibinfo{year}{2003}), \eprint{hep-ph/0301124}.

\bibitem[{\citenamefont{Abbott et~al.}(2016{\natexlab{a}})}]{Abbott:2016blz}
\bibinfo{author}{\bibfnamefont{B.~P.} \bibnamefont{Abbott}}
  \bibnamefont{et~al.} (\bibinfo{collaboration}{Virgo, LIGO Scientific}),
  \bibinfo{journal}{Phys. Rev. Lett.} \textbf{\bibinfo{volume}{116}},
  \bibinfo{pages}{061102} (\bibinfo{year}{2016}{\natexlab{a}}),
  \eprint{1602.03837}.

\bibitem[{\citenamefont{Abbott et~al.}(2016{\natexlab{b}})}]{Abbott:2016nmj}
\bibinfo{author}{\bibfnamefont{B.~P.} \bibnamefont{Abbott}}
  \bibnamefont{et~al.} (\bibinfo{collaboration}{Virgo, LIGO Scientific}),
  \bibinfo{journal}{Phys. Rev. Lett.} \textbf{\bibinfo{volume}{116}},
  \bibinfo{pages}{241103} (\bibinfo{year}{2016}{\natexlab{b}}),
  \eprint{1606.04855}.

\bibitem[{\citenamefont{Abbott et~al.}(2017{\natexlab{a}})}]{Abbott:2017vtc}
\bibinfo{author}{\bibfnamefont{B.~P.} \bibnamefont{Abbott}}
  \bibnamefont{et~al.} (\bibinfo{collaboration}{VIRGO, LIGO Scientific}),
  \bibinfo{journal}{Phys. Rev. Lett.} \textbf{\bibinfo{volume}{118}},
  \bibinfo{pages}{221101} (\bibinfo{year}{2017}{\natexlab{a}}),
  \eprint{1706.01812}.

\bibitem[{\citenamefont{Abbott et~al.}(2017{\natexlab{b}})}]{Abbott:2017oio}
\bibinfo{author}{\bibfnamefont{B.~P.} \bibnamefont{Abbott}}
  \bibnamefont{et~al.} (\bibinfo{collaboration}{Virgo, LIGO Scientific}),
  \bibinfo{journal}{Phys. Rev. Lett.} \textbf{\bibinfo{volume}{119}},
  \bibinfo{pages}{141101} (\bibinfo{year}{2017}{\natexlab{b}}),
  \eprint{1709.09660}.

\bibitem[{\citenamefont{Abbott
  et~al.}(2017{\natexlab{c}})}]{TheLIGOScientific:2017qsa}
\bibinfo{author}{\bibfnamefont{B.}~\bibnamefont{Abbott}} \bibnamefont{et~al.}
  (\bibinfo{collaboration}{Virgo, LIGO Scientific}), \bibinfo{journal}{Phys.
  Rev. Lett.} \textbf{\bibinfo{volume}{119}}, \bibinfo{pages}{161101}
  (\bibinfo{year}{2017}{\natexlab{c}}), \eprint{1710.05832}.

\bibitem[{\citenamefont{Abbott et~al.}(2017{\natexlab{d}})}]{Abbott:2017gyy}
\bibinfo{author}{\bibfnamefont{B.~P.} \bibnamefont{Abbott}}
  \bibnamefont{et~al.} (\bibinfo{collaboration}{Virgo, LIGO Scientific}),
  \bibinfo{journal}{Astrophys. J.} \textbf{\bibinfo{volume}{851}},
  \bibinfo{pages}{L35} (\bibinfo{year}{2017}{\natexlab{d}}),
  \eprint{1711.05578}.

\bibitem[{\citenamefont{Abbott et~al.}(2017{\natexlab{e}})}]{Monitor:2017mdv}
\bibinfo{author}{\bibfnamefont{B.~P.} \bibnamefont{Abbott}}
  \bibnamefont{et~al.} (\bibinfo{collaboration}{Virgo, Fermi-GBM, INTEGRAL,
  LIGO Scientific}), \bibinfo{journal}{Astrophys. J.}
  \textbf{\bibinfo{volume}{848}}, \bibinfo{pages}{L13}
  (\bibinfo{year}{2017}{\natexlab{e}}), \eprint{1710.05834}.

\bibitem[{\citenamefont{Aasi et~al.}(2015)}]{TheLIGOScientific:2014jea}
\bibinfo{author}{\bibfnamefont{J.}~\bibnamefont{Aasi}} \bibnamefont{et~al.}
  (\bibinfo{collaboration}{LIGO Scientific}), \bibinfo{journal}{Class. Quant.
  Grav.} \textbf{\bibinfo{volume}{32}}, \bibinfo{pages}{074001}
  (\bibinfo{year}{2015}), \eprint{1411.4547}.

\bibitem[{\citenamefont{Punturo et~al.}(2010)}]{Punturo:2010zz}
\bibinfo{author}{\bibfnamefont{M.}~\bibnamefont{Punturo}} \bibnamefont{et~al.},
  \bibinfo{journal}{Class. Quant. Grav.} \textbf{\bibinfo{volume}{27}},
  \bibinfo{pages}{194002} (\bibinfo{year}{2010}).

\bibitem[{\citenamefont{Danzmann}(1996)}]{Danzmann:1996da}
\bibinfo{author}{\bibfnamefont{K.}~\bibnamefont{Danzmann}},
  \bibinfo{journal}{Class. Quant. Grav.} \textbf{\bibinfo{volume}{13}},
  \bibinfo{pages}{A247} (\bibinfo{year}{1996}).

\bibitem[{\citenamefont{Mirshekari et~al.}(2012)\citenamefont{Mirshekari,
  Yunes, and Will}}]{Mirshekari:2011yq}
\bibinfo{author}{\bibfnamefont{S.}~\bibnamefont{Mirshekari}},
  \bibinfo{author}{\bibfnamefont{N.}~\bibnamefont{Yunes}}, \bibnamefont{and}
  \bibinfo{author}{\bibfnamefont{C.~M.} \bibnamefont{Will}},
  \bibinfo{journal}{Phys. Rev.} \textbf{\bibinfo{volume}{D85}},
  \bibinfo{pages}{024041} (\bibinfo{year}{2012}), \eprint{1110.2720}.

\bibitem[{\citenamefont{Yunes et~al.}(2016)\citenamefont{Yunes, Yagi, and
  Pretorius}}]{Yunes:2016jcc}
\bibinfo{author}{\bibfnamefont{N.}~\bibnamefont{Yunes}},
  \bibinfo{author}{\bibfnamefont{K.}~\bibnamefont{Yagi}}, \bibnamefont{and}
  \bibinfo{author}{\bibfnamefont{F.}~\bibnamefont{Pretorius}},
  \bibinfo{journal}{Phys. Rev.} \textbf{\bibinfo{volume}{D94}},
  \bibinfo{pages}{084002} (\bibinfo{year}{2016}), \eprint{1603.08955}.

\bibitem[{\citenamefont{Keppel and Ajith}(2010)}]{Keppel:2010qu}
\bibinfo{author}{\bibfnamefont{D.}~\bibnamefont{Keppel}} \bibnamefont{and}
  \bibinfo{author}{\bibfnamefont{P.}~\bibnamefont{Ajith}},
  \bibinfo{journal}{Phys. Rev.} \textbf{\bibinfo{volume}{D82}},
  \bibinfo{pages}{122001} (\bibinfo{year}{2010}), \eprint{1004.0284}.

\bibitem[{\citenamefont{Bailey and Havert}(2017)}]{Bailey:2017lbo}
\bibinfo{author}{\bibfnamefont{Q.~G.} \bibnamefont{Bailey}} \bibnamefont{and}
  \bibinfo{author}{\bibfnamefont{D.}~\bibnamefont{Havert}},
  \bibinfo{journal}{Phys. Rev.} \textbf{\bibinfo{volume}{D96}},
  \bibinfo{pages}{064035} (\bibinfo{year}{2017}), \eprint{1706.10157}.

\bibitem[{\citenamefont{Tso et~al.}(2017)\citenamefont{Tso, Isi, Chen, and
  Stein}}]{Tso:2016mvv}
\bibinfo{author}{\bibfnamefont{R.}~\bibnamefont{Tso}},
  \bibinfo{author}{\bibfnamefont{M.}~\bibnamefont{Isi}},
  \bibinfo{author}{\bibfnamefont{Y.}~\bibnamefont{Chen}}, \bibnamefont{and}
  \bibinfo{author}{\bibfnamefont{L.}~\bibnamefont{Stein}}, in
  \emph{\bibinfo{booktitle}{{Proceedings, 7th Meeting on CPT and Lorentz
  Symmetry (CPT 16): Bloomington, Indiana, USA, June 20-24, 2016}}}
  (\bibinfo{year}{2017}), pp. \bibinfo{pages}{205--208}, \eprint{1608.01284},
  \urlprefix\url{https://inspirehep.net/record/1479158/files/arXiv:1608.01284.pdf}.

\bibitem[{\citenamefont{Ade et~al.}(2016)}]{Ade:2015xua}
\bibinfo{author}{\bibfnamefont{P.~A.~R.} \bibnamefont{Ade}}
  \bibnamefont{et~al.} (\bibinfo{collaboration}{Planck Collaboration}),
  \bibinfo{journal}{Astron. Astrophys.} \textbf{\bibinfo{volume}{594}},
  \bibinfo{pages}{A13} (\bibinfo{year}{2016}), \eprint{1502.01589}.

\bibitem[{\citenamefont{Ajith et~al.}(2011)}]{Ajith:2009bn}
\bibinfo{author}{\bibfnamefont{P.}~\bibnamefont{Ajith}} \bibnamefont{et~al.},
  \bibinfo{journal}{Phys. Rev. Lett.} \textbf{\bibinfo{volume}{106}},
  \bibinfo{pages}{241101} (\bibinfo{year}{2011}), \eprint{0909.2867}.

\bibitem[{\citenamefont{Yunes et~al.}(2010)\citenamefont{Yunes, O'Shaughnessy,
  Owen, and Alexander}}]{Yunes:2010yf}
\bibinfo{author}{\bibfnamefont{N.}~\bibnamefont{Yunes}},
  \bibinfo{author}{\bibfnamefont{R.}~\bibnamefont{O'Shaughnessy}},
  \bibinfo{author}{\bibfnamefont{B.~J.} \bibnamefont{Owen}}, \bibnamefont{and}
  \bibinfo{author}{\bibfnamefont{S.}~\bibnamefont{Alexander}},
  \bibinfo{journal}{Phys. Rev.} \textbf{\bibinfo{volume}{D82}},
  \bibinfo{pages}{064017} (\bibinfo{year}{2010}), \eprint{1005.3310}.

\bibitem[{\citenamefont{Sathyaprakash and Schutz}(2009)}]{Sathyaprakash:2009xs}
\bibinfo{author}{\bibfnamefont{B.~S.} \bibnamefont{Sathyaprakash}}
  \bibnamefont{and} \bibinfo{author}{\bibfnamefont{B.~F.}
  \bibnamefont{Schutz}}, \bibinfo{journal}{Living Rev. Rel.}
  \textbf{\bibinfo{volume}{12}}, \bibinfo{pages}{2} (\bibinfo{year}{2009}),
  \eprint{0903.0338}.

\bibitem[{\citenamefont{Li}(2013)}]{Li:2013lza}
\bibinfo{author}{\bibfnamefont{T.~G.~F.} \bibnamefont{Li}}, Ph.D. thesis,
  \bibinfo{school}{Vrije U., Amsterdam} (\bibinfo{year}{2013}),
  \urlprefix\url{http://inspirehep.net/record/1266133/files/thesis\_T\_G\_F\_Li.pdf}.

\bibitem[{\citenamefont{Zhao et~al.}(2011)\citenamefont{Zhao, Van Den~Broeck,
  Baskaran, and Li}}]{Zhao:2010sz}
\bibinfo{author}{\bibfnamefont{W.}~\bibnamefont{Zhao}},
  \bibinfo{author}{\bibfnamefont{C.}~\bibnamefont{Van Den~Broeck}},
  \bibinfo{author}{\bibfnamefont{D.}~\bibnamefont{Baskaran}}, \bibnamefont{and}
  \bibinfo{author}{\bibfnamefont{T.~G.~F.} \bibnamefont{Li}},
  \bibinfo{journal}{Phys. Rev.} \textbf{\bibinfo{volume}{D83}},
  \bibinfo{pages}{023005} (\bibinfo{year}{2011}), \eprint{1009.0206}.

\bibitem[{\citenamefont{Klein et~al.}(2016)}]{Klein:2015hvg}
\bibinfo{author}{\bibfnamefont{A.}~\bibnamefont{Klein}} \bibnamefont{et~al.},
  \bibinfo{journal}{Phys. Rev.} \textbf{\bibinfo{volume}{D93}},
  \bibinfo{pages}{024003} (\bibinfo{year}{2016}), \eprint{1511.05581}.

\bibitem[{\citenamefont{Babak et~al.}(2017)\citenamefont{Babak, Gair, Sesana,
  Barausse, Sopuerta, Berry, Berti, Amaro-Seoane, Petiteau, and
  Klein}}]{Babak:2017tow}
\bibinfo{author}{\bibfnamefont{S.}~\bibnamefont{Babak}},
  \bibinfo{author}{\bibfnamefont{J.}~\bibnamefont{Gair}},
  \bibinfo{author}{\bibfnamefont{A.}~\bibnamefont{Sesana}},
  \bibinfo{author}{\bibfnamefont{E.}~\bibnamefont{Barausse}},
  \bibinfo{author}{\bibfnamefont{C.~F.} \bibnamefont{Sopuerta}},
  \bibinfo{author}{\bibfnamefont{C.~P.~L.} \bibnamefont{Berry}},
  \bibinfo{author}{\bibfnamefont{E.}~\bibnamefont{Berti}},
  \bibinfo{author}{\bibfnamefont{P.}~\bibnamefont{Amaro-Seoane}},
  \bibinfo{author}{\bibfnamefont{A.}~\bibnamefont{Petiteau}}, \bibnamefont{and}
  \bibinfo{author}{\bibfnamefont{A.}~\bibnamefont{Klein}},
  \bibinfo{journal}{Phys. Rev.} \textbf{\bibinfo{volume}{D95}},
  \bibinfo{pages}{103012} (\bibinfo{year}{2017}), \eprint{1703.09722}.

\bibitem[{\citenamefont{Finn and Chernoff}(1993)}]{Finn:1992xs}
\bibinfo{author}{\bibfnamefont{L.~S.} \bibnamefont{Finn}} \bibnamefont{and}
  \bibinfo{author}{\bibfnamefont{D.~F.} \bibnamefont{Chernoff}},
  \bibinfo{journal}{Phys. Rev.} \textbf{\bibinfo{volume}{D47}},
  \bibinfo{pages}{2198} (\bibinfo{year}{1993}), \eprint{gr-qc/9301003}.

\bibitem[{\citenamefont{Cutler and Flanagan}(1994)}]{Cutler:1994ys}
\bibinfo{author}{\bibfnamefont{C.}~\bibnamefont{Cutler}} \bibnamefont{and}
  \bibinfo{author}{\bibfnamefont{E.~E.} \bibnamefont{Flanagan}},
  \bibinfo{journal}{Phys. Rev.} \textbf{\bibinfo{volume}{D49}},
  \bibinfo{pages}{2658} (\bibinfo{year}{1994}), \eprint{gr-qc/9402014}.

\bibitem[{\citenamefont{Poisson and Will}(1995)}]{Poisson:1995ef}
\bibinfo{author}{\bibfnamefont{E.}~\bibnamefont{Poisson}} \bibnamefont{and}
  \bibinfo{author}{\bibfnamefont{C.~M.} \bibnamefont{Will}},
  \bibinfo{journal}{Phys. Rev.} \textbf{\bibinfo{volume}{D52}},
  \bibinfo{pages}{848} (\bibinfo{year}{1995}), \eprint{gr-qc/9502040}.

\bibitem[{\citenamefont{Ajith}(2011)}]{Ajith:2011ec}
\bibinfo{author}{\bibfnamefont{P.}~\bibnamefont{Ajith}},
  \bibinfo{journal}{Phys. Rev.} \textbf{\bibinfo{volume}{D84}},
  \bibinfo{pages}{084037} (\bibinfo{year}{2011}), \eprint{1107.1267}.

\bibitem[{\citenamefont{Mishra et~al.}(2010)\citenamefont{Mishra, Arun, Iyer,
  and Sathyaprakash}}]{Mishra:2010tp}
\bibinfo{author}{\bibfnamefont{C.~K.} \bibnamefont{Mishra}},
  \bibinfo{author}{\bibfnamefont{K.~G.} \bibnamefont{Arun}},
  \bibinfo{author}{\bibfnamefont{B.~R.} \bibnamefont{Iyer}}, \bibnamefont{and}
  \bibinfo{author}{\bibfnamefont{B.~S.} \bibnamefont{Sathyaprakash}},
  \bibinfo{journal}{Phys. Rev.} \textbf{\bibinfo{volume}{D82}},
  \bibinfo{pages}{064010} (\bibinfo{year}{2010}), \eprint{1005.0304}.

\bibitem[{\citenamefont{Cram{\'e}r}(1999)}]{cramer1999mathematical}
\bibinfo{author}{\bibfnamefont{H.}~\bibnamefont{Cram{\'e}r}},
  \emph{\bibinfo{title}{Mathematical methods of statistics}},
  vol.~\bibinfo{volume}{9} (\bibinfo{publisher}{Princeton university press},
  \bibinfo{year}{1999}).

\bibitem[{\citenamefont{Rao}(1945)}]{Rao:1945}
\bibinfo{author}{\bibfnamefont{C.~R.} \bibnamefont{Rao}},
  \bibinfo{journal}{Bullet. Calcutta Math. Soc.} \textbf{\bibinfo{volume}{37}},
  \bibinfo{pages}{81} (\bibinfo{year}{1945}).

\bibitem[{\citenamefont{Yunes and Pretorius}(2009)}]{Yunes:2009ke}
\bibinfo{author}{\bibfnamefont{N.}~\bibnamefont{Yunes}} \bibnamefont{and}
  \bibinfo{author}{\bibfnamefont{F.}~\bibnamefont{Pretorius}},
  \bibinfo{journal}{Phys. Rev.} \textbf{\bibinfo{volume}{D80}},
  \bibinfo{pages}{122003} (\bibinfo{year}{2009}), \eprint{0909.3328}.

\bibitem[{\citenamefont{Arun et~al.}(2006)\citenamefont{Arun, Iyer, Qusailah,
  and Sathyaprakash}}]{Arun:2006hn}
\bibinfo{author}{\bibfnamefont{K.~G.} \bibnamefont{Arun}},
  \bibinfo{author}{\bibfnamefont{B.~R.} \bibnamefont{Iyer}},
  \bibinfo{author}{\bibfnamefont{M.~S.~S.} \bibnamefont{Qusailah}},
  \bibnamefont{and} \bibinfo{author}{\bibfnamefont{B.~S.}
  \bibnamefont{Sathyaprakash}}, \bibinfo{journal}{Phys. Rev.}
  \textbf{\bibinfo{volume}{D74}}, \bibinfo{pages}{024006}
  (\bibinfo{year}{2006}), \eprint{gr-qc/0604067}.

\bibitem[{\citenamefont{Agathos et~al.}(2014)\citenamefont{Agathos, Del~Pozzo,
  Li, Van Den~Broeck, Veitch, and Vitale}}]{Agathos:2013upa}
\bibinfo{author}{\bibfnamefont{M.}~\bibnamefont{Agathos}},
  \bibinfo{author}{\bibfnamefont{W.}~\bibnamefont{Del~Pozzo}},
  \bibinfo{author}{\bibfnamefont{T.~G.~F.} \bibnamefont{Li}},
  \bibinfo{author}{\bibfnamefont{C.}~\bibnamefont{Van Den~Broeck}},
  \bibinfo{author}{\bibfnamefont{J.}~\bibnamefont{Veitch}}, \bibnamefont{and}
  \bibinfo{author}{\bibfnamefont{S.}~\bibnamefont{Vitale}},
  \bibinfo{journal}{Phys. Rev.} \textbf{\bibinfo{volume}{D89}},
  \bibinfo{pages}{082001} (\bibinfo{year}{2014}), \eprint{1311.0420}.

\bibitem[{\citenamefont{Schutz}(2011)}]{Schutz:2011tw}
\bibinfo{author}{\bibfnamefont{B.~F.} \bibnamefont{Schutz}},
  \bibinfo{journal}{Class. Quant. Grav.} \textbf{\bibinfo{volume}{28}},
  \bibinfo{pages}{125023} (\bibinfo{year}{2011}), \eprint{1102.5421}.

\bibitem[{\citenamefont{Samajdar and Arun}(2017)}]{Samajdar:2017mka}
\bibinfo{author}{\bibfnamefont{A.}~\bibnamefont{Samajdar}} \bibnamefont{and}
  \bibinfo{author}{\bibfnamefont{K.~G.} \bibnamefont{Arun}},
  \bibinfo{journal}{Phys. Rev.} \textbf{\bibinfo{volume}{D96}},
  \bibinfo{pages}{104027} (\bibinfo{year}{2017}), \eprint{1708.00671}.

\bibitem[{\citenamefont{Thrane and Romano}(2013)}]{Thrane:2013oya}
\bibinfo{author}{\bibfnamefont{E.}~\bibnamefont{Thrane}} \bibnamefont{and}
  \bibinfo{author}{\bibfnamefont{J.~D.} \bibnamefont{Romano}},
  \bibinfo{journal}{Phys. Rev.} \textbf{\bibinfo{volume}{D88}},
  \bibinfo{pages}{124032} (\bibinfo{year}{2013}), \eprint{1310.5300}.

\end{thebibliography}
\end{document}